\documentclass[a4paper,11pt]{article}
\usepackage{pos}
\usepackage{amsmath,amssymb}

\title{Multimessenger astronomy driven by high-energy neutrinos}
 \ShortTitle{Neutrino Multimessenger}

\author*[a]{Shigeru Yoshida}

\affiliation[a]{International Center for Hadron Astrophysics, Chiba University, Chiba 263-8522, Japan}

\emailAdd{syoshida@hepburn.s.chiba-u.ac.jp}

\abstract{The possible connection between high energy neutrinos in the energy region above 100 TeV
  and ultrahigh energy cosmic rays (UHECRs) at energies above $10^{19}$~eV motivates multi-messenger observation
  approaches involving neutrinos and the multi-wavelength electro-magnetic (EMG) signals. We have constructed
  a generic unification scheme to model the neutrino and UHECR common sources. Finding the allowed space
  of the parameters on the source characteristics allows a case study to evaluate the likelihood
  of each of the known source classes being such unified sources. The likely source candidates are transient
  or flaring objects mainly in optical and X-ray bands. We propose the two feasible strategies to identify these sources.
  One is to introduce a sub-threshold triggering in a wide field of view X-ray observatory
  for following up neutrino detections, and the other is
  to search for EMG counterparts associated with detections of multiple neutrino events coming from the same direction
  within a time scale of $\lesssim 30$~days. Sources with a total neutrino emission energy greater than
  $\sim 10^{51}$~erg are accessible with the present or near-future high energy neutrino observation facilities
  collaborating with X-rays and optical telescopes currently in operation. The neutrino-driven multi-messenger
  observations provide a smoking gun to probe the hadronic emission sources we would not be able to find otherwise.}

\ConferenceLogo{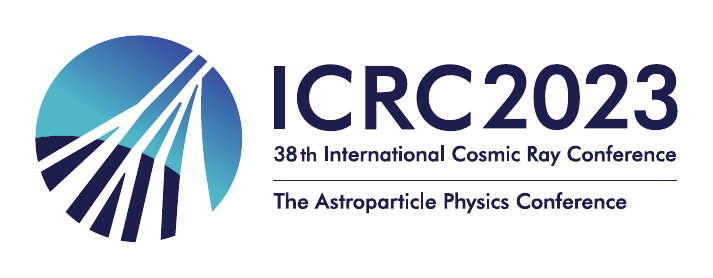}

\FullConference{%
38th International Cosmic Ray Conference (ICRC2023)\\
  26 July - 3 August, 2023\\
  Nagoya, Japan}

\begin{document}
\maketitle

\section{Introduction}

\begin{figure*}[t]
\begin{center}
\includegraphics[width=0.6\textwidth]{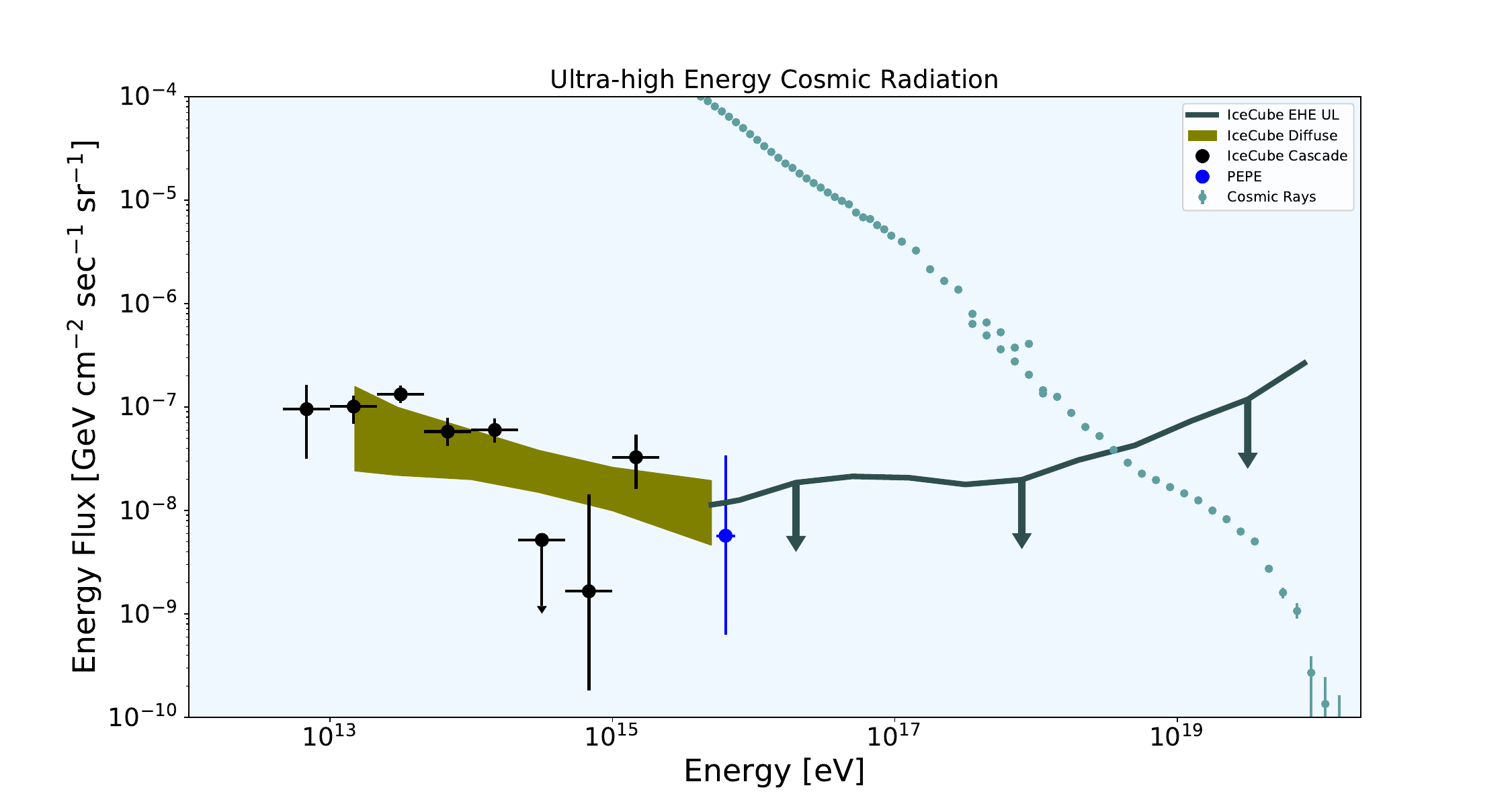}
\caption{The energy fluxes of UHE cosmic background radiations. The small data points
  represent the UHECR fluxes measured by PAO~\cite{Aab:2016zth}. The rest of the data shows the neutrino energy fluxes and
  the upper limits measured with IceCube. The thick black data points were obtained
  by the neutrino-induced cascade measurements~\cite{Aartsen:2020aqd}. The shaded region indicates the flux space
  consistent with the $\nu_\mu$ induced track measurements~\cite{IceCube:2021uhz}. The blue data point shows the flux
  of the 6 PeV-energy $\bar{\nu_e}$ event estimated by the dedicated search channel (PEPE)~\cite{IceCube:2021rpz}.
  The thick line with arrows indicates the differential upper limit obtained by
  the Extremely-high Energy (EHE) neutrino search~\cite{Aartsen:2018vtx}.
  The neutrino fluxes are the all-flavor-sum fluxes $E_\nu^2 \Phi_{\nu_e + \nu_\mu + \nu_\tau}$.
 }
\label{fig:energy_fluxes}
\end{center}
\end{figure*}

The cosmic background radiation in ultra-high energy (UHE) sky at $\gg$~TeV is formed
by cosmic rays and neutrinos. The precise measurements of
ultrahigh-energy cosmic rays (UHECRs) by Pierre Auger Observatory (PAO)
with high statistics have now revealed the detailed structure of their energy spectrum~\cite{Aab:2016zth}.
The IceCube Neutrino Observatory has discovered~\cite{Aartsen:2013bka, Aartsen:2013jdh} and measured the high energy neutrino radiation
in the UHE sky~\cite{Aartsen:2020aqd, IceCube:2021uhz}, realizing the observation window of the penetrating messengers
to study the UHE emissions. As shown in Fig.~\ref{fig:energy_fluxes}, we have found that
the observed energy flux of high-energy neutrinos above $~\sim 100$~TeV
is comparable to that of UHECRs at $\gtrsim 10^{19}$~eV.
It suggests that the neutrino radiation may originate in the same astronomical objects
to radiate UHECRs we have been detecting. It may be plausible that
the UHE cosmic radiation can be understood in a common unified scheme.

We have built the unification generic model to account for
the observed neutrinos at energies greater than $\sim 100$~TeV and UHECRs,
based on the photo-hadronic interaction framework~\cite{Yoshida:2020div}.
This modeling enables us
to evaluate if a given class of astronomical objects are qualified as
the possible common origin of UHECRs and neutrinos. It is a viable tool
to probe the UHECR origin with the multi-messenger observations.

In this article, we discuss the possibilities of the UHECR-neutrino common sources
for a broad spectrum of the astronomical objects classes with the generic unification model.
We then discuss how we can identify these sources.
As the viable source candidates are transients, we argue that
neutrino follow-up observations in optical and X-ray bands are
feasible methods to find the sources of hadronic emissions.
We propose practical strategies to pin down neutrino sources
that could not be identified by neutrino observations alone.
Finally, we conclude that multi-messenger observations with neutrinos, optical,
and X-ray photons pioneer in the field of high energy astrophysics
with the presently and soon-to-be available facilities.

\section{A generic unification model}

We constructed the generic framework to describe a unified origin of UHECRs
and high energy neutrinos with the parameterization less dependent on
the details of the source environment and the model-dependent micro-physics
processes~\cite{Yoshida:2020div}. Our goal here is to derive generic and model-independent constraints
on characteristics of the possible UHECR-neutrino common emitters.
The resultant constraints are considered
to be conservative and weaker than those obtained by the model-dependent
arguments applied to each of the specific classes of astronomical objects,
but the results based on our generic framework are universal and robust
as long as the UHE particle emission mechanism
is mostly determined in a simple setup with physics processes well approximated
by a first leading order effect. We introduce the Occam's razor principle
here to judge whether a given astronomical object class can be the unified origin.
In this sense, our arguments provide a guidance for astronomers
to conduct the multi-messenger observations.

We made the following assumptions to build our generic unification framework.

\begin{description}
\item[One zone]\mbox{}\\
  The cosmic ray accelerations and their interactions to produce secondary neutrinos
  occur in the same place.
\item[Escape from sources]\mbox{}\\
  The energy spectrum of accelerated UHECRs running away from the acceleration zone
  is not drastically distorted in their escape process. This assumption may affect the UHECR energetics
  condition we will discuss below. Considering the uncertainties of the UHECR escape process,
  we use the measured UHECR intensity as the upper limit to constrain the source luminosity
  rather than to fit it with the spectrum calculated by the generic model.
\item[Optically thin environment]\mbox{}\\
  The sources are optically thin for UHE protons interacting with photons, and their emission is directly
  observed without absorption. This assumption is valid for photo-hadronic scenarios with one zone modelling.
\item[Photon spectrum]\mbox{}\\
  The spectrum of photons interacting with UHECRs to produce neutrinos is described by a power-law.
  However we note that the thermal photon yield can also be reasonably approximated by a power-law form
  in the energy range relevant to the high energy neutrino emission within a factor of two allowance.
\item[Cosmological evolution]\mbox{}\\
  The UHECR-neutrino common sources follow the cosmological evolution tracing the star formation rate (SFR)
  or any other similar evolutions.
  However, we parameterize the evolutions to estimate the conditions for
  the different evolution cases when necessary.
\end{description}

\subsection{The source modelling}

The power of the unified sources is gauged by the bolometric photon luminosity
\begin{equation}
L'_{\gamma} = 4\pi R^2c\int\limits_{\varepsilon_{\gamma}^{\rm min}}^{\varepsilon_{\gamma}^{\rm max}}  \frac{dn_\gamma}{d\varepsilon'_\gamma}\varepsilon'_\gamma d\varepsilon'_\gamma,
 \label{eq:photon_energy_conversion}
\end{equation}
where $R$ is the distance of the UHECR acceleration and emission site from the source center.
Primed (’) characters represent quantities measured
in the rest frame of plasma with the Lorentz bulk factor $\Gamma$.
The photon density spectrum follows a power-law form as
\begin{equation}
\frac{dn_\gamma}{d\varepsilon'_\gamma}= \frac{K'_\gamma}{\varepsilon'_{\gamma0}}\left(\frac{\varepsilon'_\gamma}{\varepsilon'_{\gamma0}}\right)^{-\alpha_\gamma},
 \label{eq:target_photon}
\end{equation}
where $\varepsilon'_{\gamma0}$ is the reference energy in the engine frame,
and it is associated with the representative energy of UHECRs $\varepsilon_{p0}^\Delta$
by
\begin{equation}
\varepsilon_{\gamma0}= \frac{(s_R-m_p^2)}{4}\frac{\Gamma^2}{\varepsilon_{p0}^\Delta}
\label{eq:photon_energy_ref}
\end{equation}
where $s_R\approx 1.47$~GeV$^2$ is the Mandelstam variable at the $\Delta$ resonance
in the photopion production. The representative CR energy $\varepsilon_{p0}^\Delta$
is set to be 10~PeV in the present formulation, as this energy range of cosmic ray protons
should produce the PeV-energy neutrinos IceCube has detected. The spectrum of UHECRs emitted
from the sources is assumed to follow a power-law with index of $\alpha_{\rm CR}$.


The bolometric luminosity of UHECRs at energies above $\varepsilon_{p0}^\Delta=10$~PeV
is connected to $L_\gamma$ via the CR loading factor $\xi_{\rm CR}$ as
$L_{\rm CR}\approx \xi_{\rm CR}L_\gamma = \xi_{\rm CR}L'_\gamma \Gamma^2$.

The neutrino luminosity with respect to a given $L_{\rm CR}$
is determined by the $p\gamma$ interaction optical depth,
an average number of the interaction times before cosmic ray protons
escape from the interaction site. It is approximately given by~\cite{Yoshida:2020div}

\begin{eqnarray}
\tau_{p\gamma}(\varepsilon_i) &\approx & \tau_{p\gamma0} {\left(\frac{\varepsilon_i}{{\tilde{\varepsilon}_{p0}^{\Delta}}}\right)}^{\alpha_\gamma-1} \nonumber \\
&\approx& \left[\frac{2}{1+\alpha_\gamma}\frac{K'_{\gamma}R}{\Gamma} \int ds \frac{\sigma_{p\gamma}(s)}{s-m_p^2}\right]{\left(\frac{\varepsilon_i}{{\tilde{\varepsilon}_{p0}^{\Delta}}}\right)}^{\alpha_\gamma-1} \label{eq:optical_depth} \\
&\approx & \frac{B'}{\Gamma^2}\sqrt{\frac{L'_{\gamma}}{\xi_B}} C(\alpha_\gamma,\tilde{\varepsilon}_{p0}^{\Delta})
{\left(\frac{\varepsilon_i}{{\tilde{\varepsilon}_{p0}^{\Delta}}}\right)}^{\alpha_\gamma-1}.
\label{eq:optical_depth2}
\end{eqnarray}
Proceeding from Eq.~(\ref{eq:optical_depth}) to (\ref{eq:optical_depth2}),
the explicit dependence on $R$
is eliminated by considering the energy density balance between the photon radiation $L'_\gamma$
and the magnetic energy with the B-field strength $B'$ via the equipartition parameter $\xi_B$.
The constant $C$ depends only
on the photon spectrum power-law index $\alpha_\gamma$ and the representative CR energy
 $\varepsilon_{p0}^\Delta$ and is approximately given by

\begin{eqnarray}
C(\alpha_\gamma,\tilde{\varepsilon}_{p0}^{\Delta})&\sim&2.4\times{10}^{-24}~{\rm erg}^{-1}~{\rm cm}^{3/2}~{\rm s}^{1/2}\nonumber\\
&\times&\left(\frac{2}{1+\alpha_\gamma}\right)
\left(\frac{\tilde{\varepsilon}_{p0}^{\Delta}}{10~{\rm PeV}}\right).
\end{eqnarray}

\subsection{The required source conditions}

A UHECR source must meet the following necessity conditions : The acceleration,
the escape, and the survival requirements.

In order to accelerate cosmic rays to UHE range, the acceleration time
must be faster than the dynamical time scale. It sets the lower bound
for the magnetic field as
\begin{equation}
    B'\gtrsim \frac{\varepsilon_i^{\rm max}}{eZ}\frac{\eta}{R}\approx 1.1\times 10^5 \eta \left(\frac{R}{3\times 10^{12}\ {\rm cm}}\right)^{-1}
    \left(\frac{\varepsilon_i^{\rm max}}{Z10^{11}~{\rm GeV}}\right)
    \ {\rm G},
    \label{eq:hillas_condition2}
\end{equation}
where $\eta \gtrsim 1$ is the particle acceleration efficiency term.
This condition, also known as
the Hillas condition when $\eta\rightarrow \beta^{-2}$, can be transformed
to the constraint on the target photon
luminosity $L'_\gamma$, the gauge of the source engine power in the present
generic modelling scheme,
\begin{eqnarray}
L'_\gamma &\geq& \frac{1}{2}\xi_B^{-1} c\eta^2\beta^2\left(\frac{\varepsilon_i^{\rm max}}{Z e}\right)^2  \label{eq:hillas}\\
&\simeq&1.7\times 10^{45}~\xi_B^{-1}\eta^{2}\beta^2\left(\frac{\varepsilon_i^{\rm max}}{Z10^{11}~{\rm GeV}}\right)^2\quad{\rm erg/s}
\label{eq:hillas_condition}
\end{eqnarray}

In addition, to ensure that UHECRs can leave the
sources before losing their energies, the escape time scale
must be faster than the cosmic ray energy loss time scale.
The energy loss processes consist of the $p\gamma$ photo-meson production,
Bethe-Heitler (BH) interactions, and the synchrotron cooling.
The photo-meson production time scale is essentially
counted with the $p\gamma$ optical depth, $\tau_{p\gamma}$,
in the present scheme, and any sources with $\tau_{p\gamma}(\varepsilon_i^{\rm max})\lesssim 1$
implies that the energy loss by the $p\gamma$ photo-meson production
is not a deciding factor to limit the UHECR acceleration and escape processes.
We examine this $\tau_{p\gamma}$ condition 
by estimating $\tau_{p\gamma}$ using Eq.~(\ref{eq:optical_depth2})
in Section~\ref{sec:case_study}.
As the BH process is in general important only if the photon spectrum is softer as $\alpha_\gamma\gtrsim 2$,
the UHECR escape condition is formulated as a necessity condition by requiring
the dynamical time scale faster than the synchrotron cooling time scale.
It has been found that this condition is transformed to the upper bound
of the $p\gamma$ optical depth at the cosmic ray reference energy
$\varepsilon_{p0}^\Delta = 10$~PeV, $\tau_{p\gamma0}$, as
\begin{eqnarray}
\tau_{p\gamma0} &\lesssim & 6\times 10^{-1} \frac{2}{1+\alpha_\gamma} \left(\frac{\xi_B}{0.1}\right)^{-1}{\left(\frac{A}{Z}\right)}^4{\left(\frac{\varepsilon_i^{\rm max}}{10^{11}\ {\rm GeV}}\right)}^{-1}.
\label{eq:esc_condition}
\end{eqnarray}

If the measured bulk of UHECRs is dominated by heavier nuclear rather than nucleon
as strongly indicated by the data obtained by PAO,
the further severe requirement must be satisfied -- The nuclei survival condition~\cite{Murase:2010gj}.
That is, we require that nuclei with $A > 1$ and $Z > 1$ are accelerated and survive.
This is possible only if the time scale of the photo-disintegration is slower than
the dynamical time scale which leads to the condition on photo-disintegration optical depth
as $\tau_{A\gamma} \lesssim A$. We found that this condition sets the upper bound
of the $p\gamma$ optical depth as~\cite{Yoshida:2020div}
\begin{eqnarray}
\tau_{p\gamma0} &\lesssim& A\frac{\int ds \frac{\sigma_{p\gamma}(s)}{s-m_p^2}}
{\int ds \frac{\sigma_{A\gamma}(s)}{s-m_A^2}}
{\left[\left(\frac{s_{\rm GDR}-m_A^2}{s_R-m_p^2}\right)
\left(\frac{\tilde{\varepsilon}^{\Delta}_{p0}}{\varepsilon_{i}^{\rm max}}\right)\right]}^{\alpha_\gamma-1} \nonumber \\
&\lesssim & 0.4~{\left(\frac{A}{56}\right)}^{0.79} = 0.2\left(\frac{A}{24}\right)^{0.79}.
\label{eq:suvival_condition}
\end{eqnarray}

\subsection{The constraints due to the UHECR and and neutrino fluxes}

\begin{figure*}[t]
\begin{center}
\includegraphics[width=0.4\textwidth]{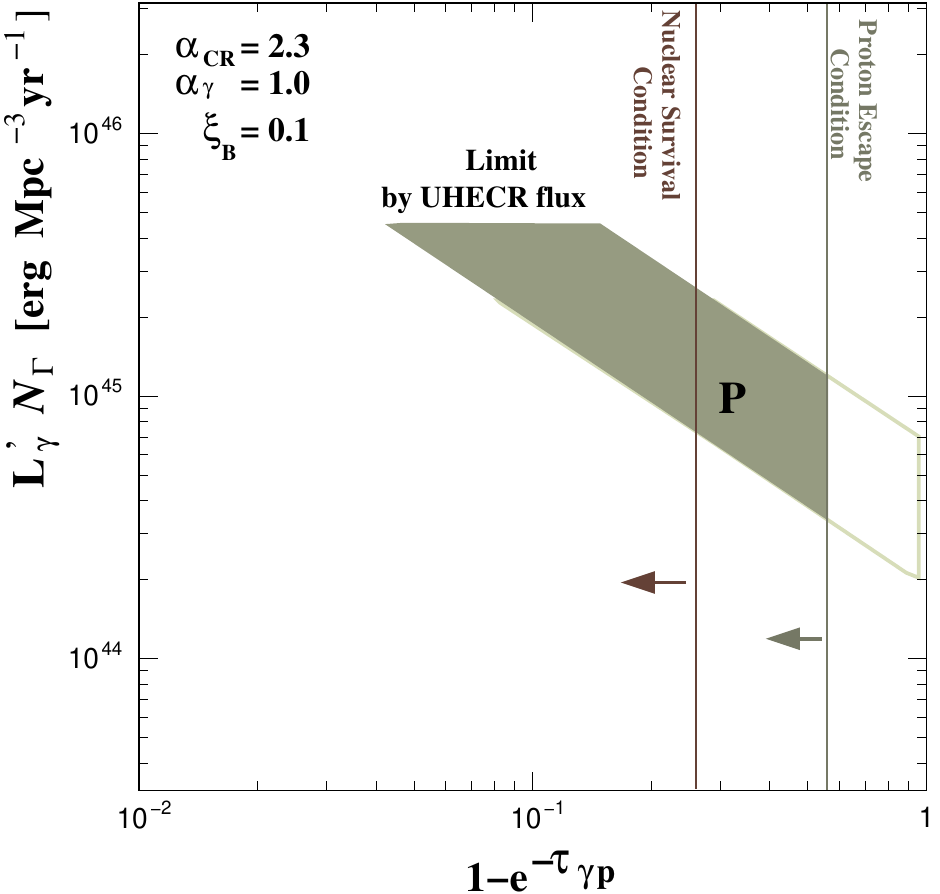}
\includegraphics[width=0.4\textwidth]{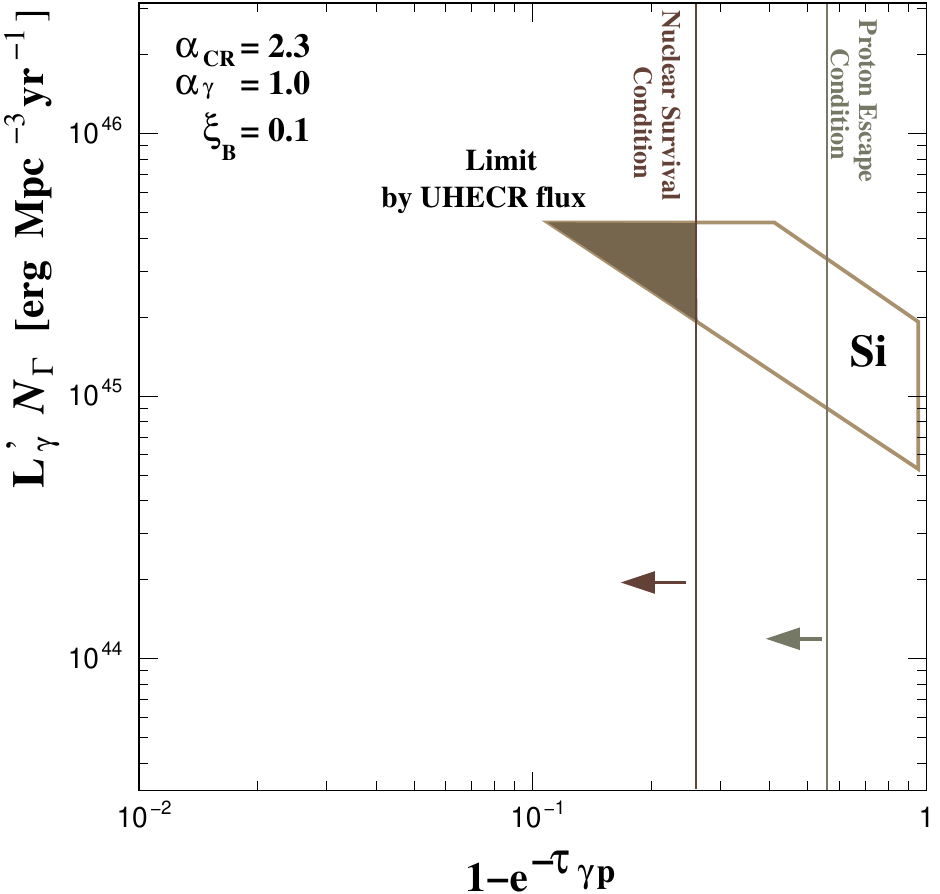}
\caption{The allowed region in the parameter space of luminosity per unit volume, $L'_\gamma\mathcal{N}_\Gamma$,
  and damping factor $\displaystyle{1-e^{-\tau_{p\gamma0}}}$~\cite{Yoshida:2020div}.
  The region enclosed by the solid lines
  displays the allowed space by the UHECR and the neutrino flux requirements.
  The shaded region represents the parameter space allowed also by considering the UHECR proton escape
  condition or the nuclear survival condition. The left panel shows the proton model while the right panel
  shows the case of primary silicon nuclei.}
\label{fig:unification_scenario_constraints}
\end{center}
\end{figure*}

UHECR sources in the unification scenario must provide both the UHECR flux and
the high energy neutrino flux that are consistent with the measurements.
The analytical formulation to calculate the spectrum of UHECRs on the Earth
and that of secondary produced neutrinos have been derived in Ref.~\cite{Yoshida:2014uka, Yoshida:2020div}
to place the constraints on the source parameters of $L'_\gamma$, $\tau_{p\gamma0}$,
and the boosted source number density $N_\Gamma\equiv n_0\xi_{\rm CR}\Gamma^2$
where $n_0$ is the comoving number density in the present epoch. Note that
$L'_\gamma N_\Gamma = \xi_{\rm CR}L_\gamma n_0 = L_{\rm CR}n_0$ is the bolometric
luminosity density of UHECRs above $\varepsilon_{p0}^\Delta = 10$~PeV.
Figure~\ref{fig:unification_scenario_constraints} shows the resultant
constraints. The region enclosed in the solid lines is the allowed parameter
space by the flux conditions. The conditions were set 
by the criteria that ensure the consistency to the neutrino measurements
with IceCube including the upper limit of flux at $\varepsilon_\nu\gtrsim 100$~PeV,
and that the UHECR flux on the earth would not exceed the integral flux above
$10^{19}$~eV measured by PAO.

We can interpret the allowed parameter space shown in Fig.~\ref{fig:unification_scenario_constraints}
from the context of the UHECR energetics and the analytical estimate of the fiducial
neutrino flux. The UHECR differential luminosity density is estimated
as~\cite{Murase:2018utn}
\begin{eqnarray}
    E_{\rm CR}\frac{dQ_{\rm CR}}{dE_{\rm CR}} &\approx& 6.3\times 10^{43} [{\rm erg\ Mpc^{-3} yr^{-1}}]\
    \left(\frac{E_{\rm CR}}{10^{19.5\ {\rm eV}}}\right)^{2-\alpha_{\rm CR}} \nonumber\\
    &\approx & \left \{ \begin{array}{lr}
    1.8\times 10^{44}\ [{\rm erg\ Mpc^{-3} yr^{-1}}] & \alpha_{\rm CR}=2.3, E_{\rm CR}=10^{18}\ {\rm eV} \\
    3.4\times 10^{44}\ [{\rm erg\ Mpc^{-3} yr^{-1}}] & \alpha_{\rm CR}=2.5, E_{\rm CR}=10^{18}\ {\rm eV}
    \end{array}
    \right.
    \label{eq:UHECR_luminosity_density}
\end{eqnarray}

As a representative case, we consider $\alpha_{\rm CR}=2.3$ hereafter.
From Eq.~(\ref{eq:UHECR_luminosity_density}), the resultant bolometric UHECR energy density
above the reference energy above $\varepsilon_{p0}^\Delta = 10$~PeV
is given by
\begin{eqnarray}
    n_0\xi_{\rm CR}L_\gamma &\approx& 13  E_{\rm CR}\frac{dQ_{\rm CR}}{dE_{\rm CR}}|_{E_{\rm CR}=10^{18}\ {\rm eV}} \nonumber \\
    &\approx& 2.3\times 10^{45}\quad {\rm erg \ Mpc^{-3} yr^{-1}},
    \label{eq:UHECR_energetics}
\end{eqnarray}
which is consistent with the allowed region of the parameter space.
This energetics condition above effectively sets the requirement of the CR loading factor
for a given $L_\Gamma$ and $n_0$ such as
\begin{equation}
    \xi_{\rm CR} \approx 0.7 \left(\frac{L_\gamma}{10^{46} {\rm erg/s}}\right)^{-1}\left(\frac{n_0}{10^{-8} {\rm Mpc^{-3}}}\right)^{-1}.
    \label{eq:cr_loading}
\end{equation}

The neutrino emissivity from a source is connected to the primary UHECR emissivity as~\cite{Murase:2015xka}
\begin{equation}
  \varepsilon_\nu^2 \frac{d\dot{N}_{\nu}}{d\varepsilon_\nu} \approx \xi_\pi <x> <y_\nu> \tau_{p\gamma}\varepsilon_{\rm CR}^2\frac{d\dot{N}_{\rm CR}}{d\varepsilon_{\rm CR}} A^{2-\alpha_{\rm CR}}
  \label{eq:nuToCr}
\end{equation}
for a hadronically thin ({\it i.e.}$\tau_{p\gamma}\lesssim 5$) source. Here
$\xi_\pi\sim 3/2$ is the average multiplicity of neutrinos from a single pion
produced by the photo-hadronic interaction, and $<y_\nu>\sim 1/4$ is the average fraction
of energy channeling into a neutrino from the secondary produced pion
and $<x>\sim 0.2$ is the average ineasticity of the $p\gamma$ collision.
Since the energy flux of the high energy cosmic background neutrinos can be approximately written
using the source emissivity $\displaystyle \varepsilon_\nu^2 d\dot{N}_{\nu}/d\varepsilon_\nu$,
we can relate the $p\gamma$ optical depth to the required bolometric UHECR luminosity density
for a given energy flux of neutrinos via Eq.~(\ref{eq:nuToCr}).
We get
\begin{eqnarray}
\tau_{p\gamma0}L'_\gamma \mathcal{N}_\Gamma=\tau_{p\gamma0}n_0\xi_{\rm CR}L_\gamma &\approx& 9.3\times 10^{43} \left(\frac{E_\nu^2\Phi_\nu}{2\times 10^{-8}\ [{\rm GeV\ cm^{-2}\ sec^{-1} sr^{-1}]}}\right)\nonumber \\
 && \times\left(\frac{\xi_z}{2.8}\right)^{-1}A^{0.3} \quad {\rm erg\ Mpc^{-3} yr^{-1}},
 \label{eq:neutrino_flux}
\end{eqnarray}
for $\alpha_{\rm CR}=2.3$. Here $\displaystyle \xi_z\equiv (1/t_{\rm H})\int dt \Psi(z)/(1+z)$ is
a dimensionless parameter that depends on the redshift evolution function $\Psi(z)$ of the sources.
This relation well represents the allowed parameter space shown in Fig.~\ref{fig:unification_scenario_constraints}.

Combining the UHECR luminosity density given by Eq.~(\ref{eq:UHECR_energetics}) with
this neutrino fiducial flux condition, Eq.~(\ref{eq:neutrino_flux}), sets the lower bound
of the $p\gamma$ optical depth, which is
\begin{eqnarray}
    \tau_{p\gamma0}&\gtrsim& 0.04 A^{0.3}\left(\frac{\xi_z}{2.8}\right)^{-1} \nonumber \\
    &\gtrsim& 0.1 \left(\frac{A}{28}\right)^{0.3}\left(\frac{\xi_z}{2.8}\right)^{-1}.
    \label{eq:neutriono_flux_requirement}
\end{eqnarray}

As the UHECR escape condition sets the {\it upper} bound of the optical depth (see Eq.~(\ref{eq:esc_condition})),
this fiducial neutrino flux requirements leads to a necessity condition of the B-field equipartition parameter.
\begin{equation}
    \xi_{\rm B} \lesssim 1.5 \left(\frac{A}{Z}\right)^4\left(\frac{\xi_z}{2.8}\right){\left(\frac{\varepsilon_i^{\rm max}}{10^{11}\ {\rm GeV}}\right)}^{-1}A^{-0.3}
    \label{eq:Bfield_bound}
\end{equation}

\subsection{The case study \label{sec:case_study}}

\begin{table}[t]
    {\tiny
  \begin{tabular}{lcccc}
    \hline\hline
    & AGN corona & BL Lac & FSRQ & Radio Gal. MAD\\
    \hline
    $\Gamma$ of the target photons & $\approx$~1 & $\approx$~10 & $\approx$~1 & $\approx$~1\\
    target photon energy: Eq.~(\ref{eq:photon_energy_ref}) &opt/UV/X-ray & X-ray & UV/Opt & UV/Opt\\
    $L_\gamma\ [{\rm erg/s}]$ & $10^{44}$ & $2\times 10^{44}$ & $4\times 10^{46}$ & $10^{41}$\\
    $n_0 \ [{\rm Mpc^{-3}}]$ & $5\times 10^{-6}$ & $3\times10^{-7}$ & $3\times 10^{-10}$ & $2\times 10^{-6}$\\
    $B'\  [{\rm G}]$ & 300 &1 & 1 & 100\\
    $\xi_{\rm B}$  & 13 & 75 & 0.1 & $4\times 10^5$\\
    \hline
    Acceleration: Eq.~(\ref{eq:hillas_condition}) & $\xi_{\rm B}\gtrsim 0.087 (\frac{Z}{14})^{-2}$ & $\xi_{\rm B}\gtrsim 4.3 (\frac{Z}{14})^{-2}$ & $\xi_{\rm B}\gtrsim 0.04 (\frac{Z}{1})^{-2}$ & $\xi_{\rm B}\gtrsim 87 (\frac{Z}{14})^{-2}$ \\
    $\tau_{p\gamma0}$ by Escape: Eq.~(\ref{eq:esc_condition}) & $\lesssim 0.005(\frac{\xi_{\rm B}}{13})^{-1}$ & $\lesssim 7.5\times 10^{-4}(\frac{\xi_B}{80})^{-1}$ & $\lesssim 0.6(\frac{\xi_B}{0.1})^{-1}$ & $\lesssim 8\times 10^{-8}(\frac{\xi_B}{4\times 10^5})^{-1}$\\
     $\tau_{p\gamma0}$ by Nuclei survival: Eq.~(\ref{eq:suvival_condition}) &$\lesssim 0.2(\frac{A}{28})^{0.79}$ &$\lesssim 0.2(\frac{A}{28})^{0.79}$ &$\lesssim 0.2(\frac{A}{28})^{0.79}$ &$\lesssim 0.2(\frac{A}{28})^{0.79}$ \\
     $\tau_{p\gamma0}$ by $\nu$ Flux: Eq.~(\ref{eq:neutriono_flux_requirement})& $\gtrsim 0.1(\frac{\xi_z}{2.8})^{-1}(\frac{A}{28})^{0.3}$ &$\gtrsim 0.1(\frac{\xi_z}{2.8})^{-1}(\frac{A}{28})^{0.3}$ & $\gtrsim 0.01(\frac{\xi_z}{8.4})^{-1}$ & $\gtrsim 0.5(\frac{\xi_z}{0.6})^{-1}(\frac{A}{28})^{0.3}$ \\
    $\xi_{\rm CR}$: Eq.~(\ref{eq:cr_loading}) &$\approx 1.2$ & $\approx 1.2$ & $\approx 5.8$ & $\approx 350$\\
    \hline\hline
    \end{tabular}
    }
    \caption{The parameters of the neutrino emission characteristics in the unification model and the constraints on
      $\xi_B$, the B-field equipartition parameter, and $\tau_{p\gamma0}$,
      the photo-hadronic optical depth at the cosmic ray reference energy of
      $\varepsilon^\Delta_{p0}=10$~PeV imposed by the conditions for UHECR-neutrino common sources.
      The various sites/population in AGN family are listed.}
    \label{table:constraints_AGN}
\end{table}

Tables~\ref{table:constraints_AGN} and \ref{table:constraints_transients}
list the various source classes together with their characteristic parameters
and the constraints on $\xi_B$ and $\tau_{p\gamma0}$ imposed by the conditions
we discussed earlier.

The AGN corona~\cite{Murase:2019vdl} is already disfavored as a common UHECR neutrino source regardless of 
any model-dependent arguments. The conditions on $\tau_{p\gamma}$ from the UHECR escape and the fiducial
neutrino flux requirements cannot hold concurrently. This is primarily because
the corona is expected to be strongly magnetized ($\xi_{\rm B}\gg 1$).
The approximate estimate using Eq.~(\ref{eq:optical_depth2}) gives
$\displaystyle  \tau_{p\gamma}\approx 100\left({B'}/{\rm 0.3 kG}\right)\left({L_\gamma}/{10^{44}{\rm erg/s}}\right)^{\frac{1}{2}}$
$\left({\xi_B}/{13}\right)^{-\frac{1}{2}}\left({\beta}/{0.02}\right)^{-1}\left({\varepsilon_i}/{\tilde{\varepsilon}_{p0}^{\Delta}}\right)^{0.8}$,
indicating that AGN corona may be a TeV-PeV neutrino source candidate, but certainly not a UHECR source
since $\tau_{p\gamma}\gg 1$  prevents hadrons from being accelerated to UHE range.

The BL Lac~\cite{2010MNRAS.402..497G} should also meet the contradicting $\tau_{p\gamma}$
conditions demanded by the UHECR escape and neutrino fiducial
flux requirements, which makes it difficult to be considered as a common source. The estimated optical depth is
$\displaystyle    \tau_{p\gamma0}\approx 4\times 10^{-6}\left({B'}/{{\rm 1 G}}\right)\left({\Gamma}/{10}\right)^{-3}$
$\left({L_\gamma}/{2\times 10^{44}{\rm erg/s}}\right)^{\frac{1}{2}}\left({\xi_B}/{80}\right)^{-\frac{1}{2}}$, which indeed suggests that BL Lacs
may be UHECR sources but too dark in neutrinos.

UHE emission from FSRQ~\cite{Murase:2014foa} is an interesting possibility.
The UHECR escape and the neutrino flux conditions can be concurrent
in principle, providing that $\tau_{p\gamma0} \sim 0.1-1$ is in the plausible range at the FSRQ system.
Using Eq.~(\ref{eq:optical_depth2}), we indeed estimate
$\displaystyle \tau_{p\gamma}\approx 1.4\left({B'}/{{\rm 1 G}}\right)\left({L_\gamma}/{4\times 10^{46}{\rm erg/s}}\right)^{\frac{1}{2}}\left({\xi_B}/{0.1}\right)^{-\frac{1}{2}} \left({\varepsilon_i}/{{\tilde{\varepsilon}_{p0}^{\Delta}}}\right)^{0.5}$.
The high photon luminosity ($L_\gamma \gtrsim 10^{46}\ {\rm erg/s}$) meets
the acceleration condition given by Eq.~(\ref{eq:hillas_condition}) for protons ($Z=1$) even if $\xi_{\rm B}\ll 1$
while the nuclei survival condition could be only barely satisfied. It implies that
FSRQs may be the common origin for UHECR protons and neutrinos, though not UHECR nuclei.
However, this hypothesis is disfavored, since the strongly evolved sources such as FSRQs are not likely
to be UHECR origins if the proton component is not negligible at the UHE range~\cite{Aartsen:2016ngq,Aab:2019auo}.
This is because the GZK cosmogenic neutrinos would have overshot the present upper limit
of neutrino flux at EeV ($10^{18}\ {\rm eV}$) range placed by IceCube and PAO.

A scenario of hadronic emissions from magnetically arrested disks (MAD) in a subclass of radio galaxies
has been proposed~\cite{Kimura:2020srn}, motivated by the GeV-TeV gamma-ray observations from nearby radio galaxies.
The cosmic ray accelerations can be plausible with the framework of radiatively inefficient
accretion flows (RIAFs). Because of the highly magnetized environment ($\xi_{\rm B}\gg 100$),
MAD may be producing UHECR nuclei though UHECR protons cannot escape.
It certainly meets the acceleration condition given by Eq.~(\ref{eq:hillas_condition})
for silicons ($Z=14$). The energetics condition, Eq.~(\ref{eq:cr_loading}), requires
$\xi_{\rm CR}\gg 10$ but this constraints can be relaxed if we expand this model to
the entire Fanaroff-Riley I galaxies by including high-excitation radio galaxies,
bringing $n_0\approx 10^{-4}\ {\rm Mpc}^{-3}$. However, it is unlikely to emit
neutrinos with the intensity measured by IceCube. The approximate estimate
using Eq.~(\ref{eq:optical_depth2}) gives 
$\displaystyle \tau_{p\gamma}\approx 2.5\times 10^{-3}\left({B'}/{{\rm 100 G}}\right)\Gamma^{-3}\left({L_\gamma}/{10^{41}{\rm erg/s}}\right)^{\frac{1}{2}}\left({\xi_B}/{3.7\times 10^5}\right)^{-\frac{1}{2}}\left({\varepsilon_i}/{{\tilde{\varepsilon}_{p0}^{\Delta}}}\right)^{1.2}$,
which is far below the fiducial neutrino flux condition, Eq.~(\ref{eq:neutriono_flux_requirement}).
This is primarily because the MAD is optically too thin,
given that the luminosity of the target photons (optical/UV) $L_\gamma\sim 10^{41}\ {\rm erg/s}$.
Note that the high magnetic field strength would cause the synchrotron cooling
of the secondary produced muons by the photo-hadronic collisions,
which suppresses neutrinos with energies higher than $\sim 10$~PeV.

\begin{table}[t]
    {\tiny
  \begin{tabular}{lcccc}
    \hline\hline
    & jetted TDE & TDE corona & Low Luminosity GRB & Engine-driven SNe\\
    &  &  &  & afterglow\\
    \hline
    $\Gamma$ of the target photons & $\approx$~10 & $\approx$~1 & $\approx$~2--10 & $\approx$~1\\
    target photon energy: Eq.~(\ref{eq:photon_energy_ref}) &X-ray & opt/UV/X-ray & X-ray & UV/Opt\\
    $L_\gamma\ [{\rm erg/s}]$ & $10^{47}$ & $3\times 10^{43}$ & $10^{47}$ & $3\times 10^{45}$\\
    $n_0 \ [{\rm Mpc^{-3}}]$ & $3\times 10^{-12}(\frac{\Delta T}{10^6 {\rm s}})$ & $4\times10^{-7}(\frac{\Delta T}{10^7 {\rm s}})$ & $3\times 10^{-11}(\frac{\Delta T}{3\times 10^3 {\rm s}})$ & $10^{-9}(\frac{\Delta T}{10^4 {\rm s}})$\\
    $B'\  [{\rm G}]$ & 500 &$10^3$ & 80 & 1.6\\
    $\xi_{\rm B}$  & 1 & 45 & 0.1 & 4\\
    \hline
    Acceleration: Eq.~(\ref{eq:hillas_condition}) & $\xi_{\rm B}\gtrsim 0.009 (\frac{Z}{14})^{-2}$ & $\xi_{\rm B}\gtrsim 0.3 (\frac{Z}{14})^{-2}$ & $\xi_{\rm B}\gtrsim 0.009 (\frac{Z}{14})^{-2}$ & $\xi_{\rm B}\gtrsim 0.3 (\frac{Z}{14})^{-2}$ \\
    $\tau_{p\gamma0}$ by Escape: Eq.~(\ref{eq:esc_condition}) & $\lesssim 0.06(\frac{\xi_{\rm B}}{1})^{-1}$ & $\lesssim 1.3\times 10^{-3}(\frac{\xi_B}{45})^{-1}$ & $\lesssim 0.6(\frac{\xi_B}{0.1})^{-1}$ & $\lesssim 0.015(\frac{\xi_B}{4})^{-1}$\\
     $\tau_{p\gamma0}$ by Nuclei survival: Eq.~(\ref{eq:suvival_condition}) &$\lesssim 0.2(\frac{A}{28})^{0.79}$ &$\lesssim 0.2(\frac{A}{28})^{0.79}$ &$\lesssim 0.2(\frac{A}{28})^{0.79}$ &$\lesssim 0.2(\frac{A}{28})^{0.79}$ \\
     $\tau_{p\gamma0}$ by $\nu$ Flux: Eq.~(\ref{eq:neutriono_flux_requirement})& $\gtrsim 0.5(\frac{\xi_z}{0.5})^{-1}(\frac{A}{28})^{0.3}$ &$\gtrsim 0.5(\frac{\xi_z}{0.5})^{-1}(\frac{A}{28})^{0.3}$ & $\gtrsim 0.1(\frac{\xi_z}{2.8})^{-1}(\frac{A}{28})^{0.3}$ & $\gtrsim 0.1(\frac{\xi_z}{2.8})^{-1}(\frac{A}{28})^{0.3}$ \\
    $\xi_{\rm CR}$: Eq.~(\ref{eq:cr_loading}) &$\approx 220$ & $\approx 5.6$ & $\approx 24$ & $\approx 24$\\
    \hline\hline
    \end{tabular}
    }
    \caption{Same as Table~\ref{table:constraints_AGN} but the various transient objects are listed.}
    \label{table:constraints_transients}
\end{table}

Powerful transient objects~(table~\ref{table:constraints_transients})
have also been discussed in the literature
for considering UHECR origins. Jetted TDEs cannot be UHECR proton sources because of
the escape requirement, but potentially UHE nuclei sources~\cite{Biehl:2017hnb}. However, the tight margin
between the nuclei survival condition and the fiducial neutrino flux requirement
needs a fine parameter tuning for qualifying this object to be the UHECR-neutrino common sources.
The approximate estimate indeed gives
$\displaystyle \tau_{p\gamma}\approx 0.34\left({B'}/{{\rm 500 G}}\right)\left({\Gamma}/{10}\right)^{-3}\left({L_\gamma}/{10^{47}{\rm erg/s}}\right)^{\frac{1}{2}}\xi_B^{-\frac{1}{2}}\left({\varepsilon_i}/{{\tilde{\varepsilon}_{p0}^{\Delta}}}\right)$
which could barely meet the both conditions, considering the unavoidable uncertainties
of the parameter values. The more serious issue is that it is more difficult to meet the energetics condition.
The resultant CR loading factor $\xi_{\rm CR}\gg 10$ raises questions about credibility of the TDE scenario.

Non-jetted TDEs are more generous objects and may alleviate the energetics issue.
The wind driven by TDE or the possible corona
may be a plausible site for cosmic-ray acceleration~\cite{Murase:2020lnu}.
The demerit is their possible optically thick environment. The TDE corona scenario expects 
$\displaystyle \tau_{p\gamma}\approx 20\left({B'}/{{\rm 1 kG}}\right)\Gamma^{-3}\left({L_\gamma}/{3\times 10^{43}{\rm erg/s}}\right)^{\frac{1}{2}}\left({\xi_B}/{45}\right)^{-\frac{1}{2}} \left({\beta}/{0.1}\right)^{-1} \left({\varepsilon_i}/{{\tilde{\varepsilon}_{p0}^{\Delta}}}\right)^{0.8}$
and the UHE nuclei cannot survive. The high magnetic field ($\sim\ {\rm kG}$) naturally sets such dense photon environment
that break down nuclei via photo-disintegration. UHECR protons cannot escape either.

Low Luminosity GRBs~\cite{Murase:2006mm} are among the most promising candidates for the unification scenario. The estimated optical depth,
$\displaystyle  \tau_{p\gamma}\approx 0.1\left({B'}/{{\rm 80 G}}\right)\left({\Gamma}/{10}\right)^{-3}\left({L_\gamma}/{10^{47}{\rm erg/s}}\right)^{\frac{1}{2}}$
$\left({\xi_B}/{0.1}\right)^{-\frac{1}{2}}\left({\varepsilon_i}/{{\tilde{\varepsilon}_{p0}^{\Delta}}}\right)^{1.2}$
exactly meets all the conditions as listed in table~\ref{table:constraints_transients}.
The CR lading factor required for
the UHECR energetics may be high, but it can be relaxed, given that the rate density
$\rho_0\sim 3\times 10^{-7}\ {\rm Mpc}^{-3} {\rm yr}^{-1}$ is still quite uncertain. 

The external acceleration during the afterglow of engine-driven SNe has also been
among the UHECR nuclei emission models~\cite{Zhang:2018agl}. As seen in Table~\ref{table:constraints_transients},
the UHECR escape condition prevents the afterglow from being the common UHECR protons and neutrino sources
regardless of model-dependent arguments. Just as in the other transients, there is the (tight) margin to meet
both the nuclei survival condition and the fiducial neutrino flux requirement. However, the estimate
using Eq.~(\ref{eq:optical_depth2}) gives 
$\displaystyle    \tau_{p\gamma}\approx 1.1\times 10^{-4}\left({B'}/{{\rm 1.6 G}}\right)\left({\Gamma}/{10}\right)^{-3}$
$\left({L_\gamma}/{3\times 10^{45}{\rm erg/s}}\right)^{\frac{1}{2}}\left({\xi_B}/{4}\right)^{-\frac{1}{2}}\left({\varepsilon_i}/{{\tilde{\varepsilon}_{p0}^{\Delta}}}\right)^{0.7}$, which does not satisfy the neutrino flux condition. This source is hadronically too thin
to explain the 100 TeV-PeV energy neutrinos measured by IceCube.

\section{Identification of UHECR sources with neutrino follow-up observations}

Many of the potential high energy neutrino/UHECR sources are transient emitters.
Low Luminosity GRBs, or jetted-TDEs are among the representative examples.
SNe with strong interaction with circumstellar material and other non-relativistic transients
may contribute to the TeV neutrino sky~\cite{Fang:2020bkm,Murase:2017pfe} although they are unlikely to be the sources of UHECRs. 
Moreover, the neutrino emissions from jets in AGNs (e.g., FSRQs)
are expected to occur in flare rather than a steady manner~\cite{Yoshida:2022wac}. Thus it is a powerful
method to search for electromagnetic (EMG) counterparts by follow-up observations triggered by
a neutrino detection in order to identify neutrino or UHECR sources.

It is straightforward to find the neutrino-EMG association with a rare type
of the objects. The GeV-energy gamma-ray blazars detected by Fermi-LAT belong
to this category. However more abundant classes of objects such as SNe or low-luminosity GRBs
would yield more frequent chance coincidences between neutrino and EMG detections,
which makes it challenging to claim robust associations.
Moreover, the optical sky is filled with many SNe irrelevant from neutrino emissions
(e.g., type Ia) and they cause significant contamination in optical follow-ups.
The longer duration of neutrino transient emissions reaching to weeks
expected from circumstellar SNe or TDEs would yield even more severe contamination
by the unrelated SNe. The simple multi-messenger strategy faces the difficulty
here.

There are two feasible solutions. The first approach is to conduct follow-up observations
in X-ray band. The X-ray sky is quieter as SNe are not luminous X-ray transients.
The drawback is that many of the neutrino source candidates we discussed above
may not be bright enough for the existing X-ray telescopes with the rapid follow-up capability.
The low luminosity GRBs are good examples. Their dim luminosity of $L_{\rm X}\sim 10^{47}\ {\rm erg/s}$
is below the regular detection threshold unless a progenitor happens to be located in
the neighborhood of our galaxy. It is, therefore, necessary to implement
a sub-threshold detection trigger on a X-ray observatory with wide field of view.
The Monitor of All-sky X-ray Image (MAXI) telescope~\cite{2009PASJ...61..999M} can meet this demand
as it is regularly monitoring all sky at a single photon counting level.
The MAXI-IceCube coincidence search with the sub-threshold detection
algorithms will be scheduled in near future.

The second approach is to search for neutrino multiplet,
two (doublet) or more neutrinos originating from the same direction
within a certain time frame~\cite{Yoshida:2022idr}.
Since only nearby sources can yield detectable neutrino multiplets,
we can rule out any distant EMG transient counterpart candidates found in a follow-up
observation triggered by a multiplet detection.
Therefore, searches for the neutrino-optical association can be performed
under less contaminated environment. It has been found that
$\sim 90$~\% of sources to produce detectable neutrino multiplets
by a $1\ {\rm km}^3$ neutrino telescope are localized within
$z\lesssim 0.15$ while the distribution of sources to yield a singlet neutrino detection
extends up to $z\gtrsim 2$~\cite{Yoshida:2022idr}. Confining neutrino sources within the local universe
realizes the following strategy of optical follow-up observations for claiming
robust neutrino-optical associations.

\begin{figure}[t]
\begin{center}
\includegraphics[width=0.6\textwidth]{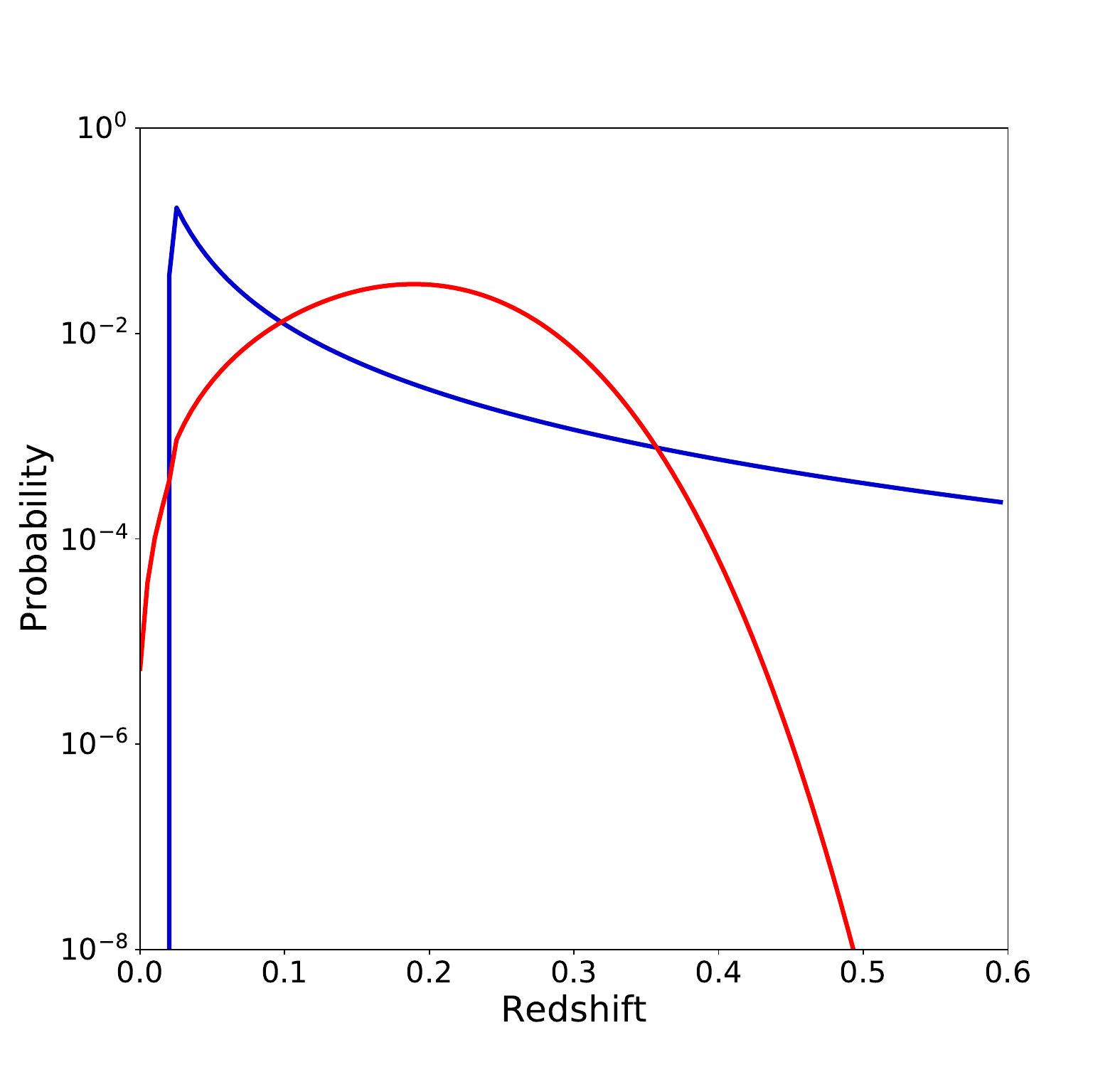}
\caption{Probability distribution of the redshift of the closest counterpart with bin size
$\Delta z = 0.005$~\cite{Yoshida:2022idr}. The bin size is chosen for illustrative purposes. 
  The blue curve represents the case of the signal hypothesis
  that the object is the neutrino source to yield the detected neutrino multiplet,
  and the red curve shows the case of
  the coincident background hypothesis, that is the chance coincident detection of
  an unassociated SN. ${\mathcal E}_\nu^{\rm fl}=1\times 10^{49}$~erg and 
$R_0=3\times 10^{-6}$~Mpc$^{-3}$~yr$^{-1}$ are assumed for the multiplet source.}
\label{fig:redshift_pdf}
\end{center}
\vspace{-1cm}
\end{figure}

We anticipate $\gtrsim$ 100 SNe in an optical follow-up observation to search for
any optical counterpart. They are mostly not associated with the neutrino detection
as mentioned earlier, 
though it is always possible that one of them can be the real counterpart.
Using the fact that sources producing neutrino multiplet are localized
in the low redshift universe, we can distinguish unassociated SNe
from the source to emit neutrino multiplet.
Among the optical transients found in a follow-up observation
triggered by a neutrino multiplet detection,
the closest object is the most likely neutrino source.
Because the expected redshift distribution of the source to yield
neutrino doublet is quite different from that of unassociated SNe,
we can judge if the closest counterpart is indeed associated with
the neutrino doublet detection in a statistical way.
Figure~\ref{fig:redshift_pdf} shows the probability distributions of
the redshift of the closest object for the two possibilities.
The pronounced difference between them can construct a test statistic
to examine which hypothesis is favored. For example,
finding an SN-like transient at $z = 0.04$ ($\approx 170\ {\rm Mpc}$) in an optical follow-up
observation leads to $\sim 2.7\sigma$
significance against the hypothesis of the incorrect coincident SN detection.

\begin{figure*}[t]
\begin{center}
\includegraphics[width=0.6\textwidth]{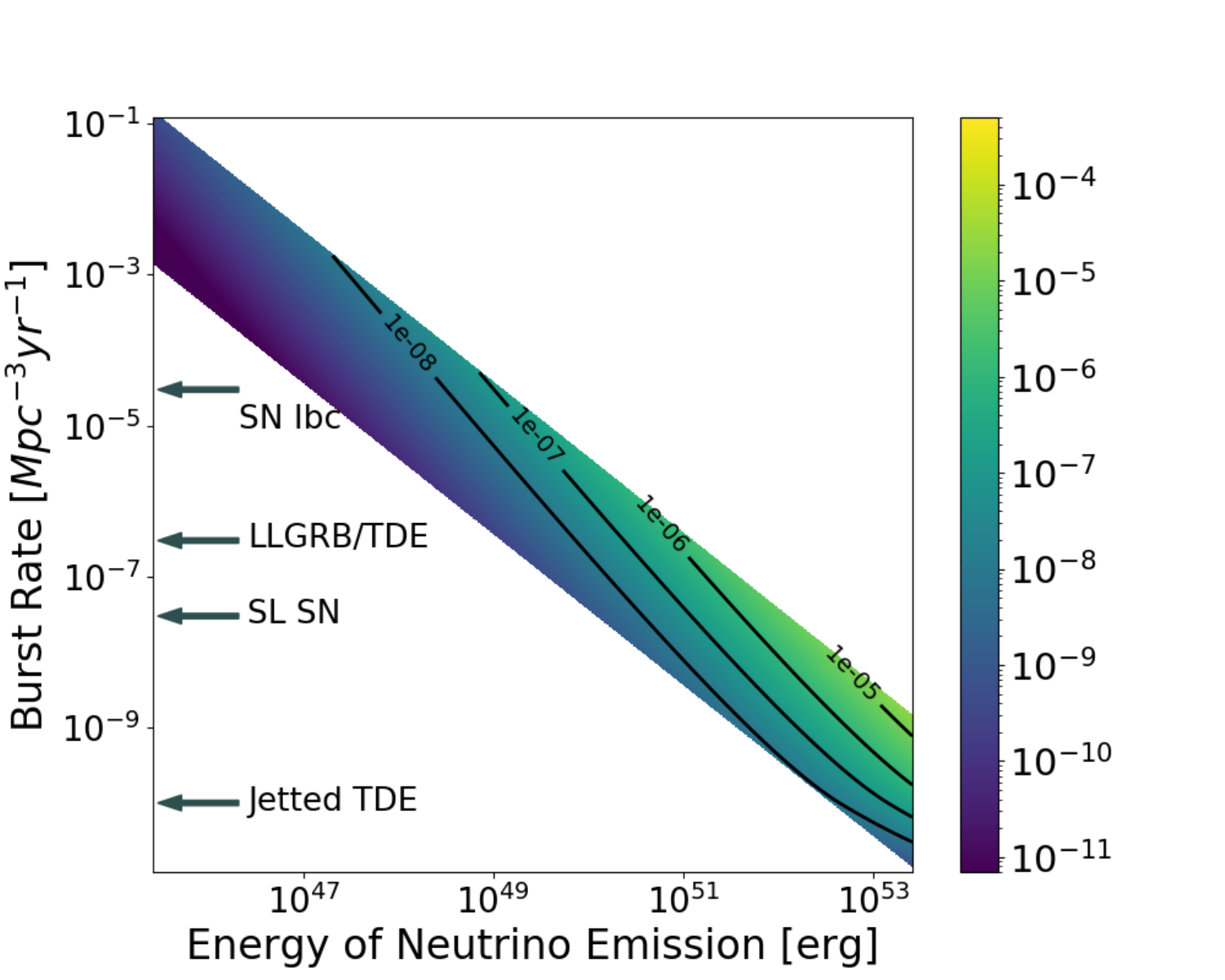}
\caption{Number of sources to yield neutrino multiplet, $N^{\rm M}_{\Delta \Omega}$,
during time $\Delta T=$~30 days 
in $\Delta \Omega= 1\ {\rm deg}^2$ of sky on the parameter space of $({\mathcal E}_\nu^{\rm fl}, R_0)$,
the output neutrino energy from a source
and the burst density rate~\cite{Yoshida:2022idr}. A criteria to suppress
the annual false alarm rate below $\sim 0.25$ for 2$\pi$ sky is applied.}
\label{fig:number_of_multiplet_s}
\end{center}
\vspace{-0.5cm}
\end{figure*}

Identifying the closest object from the numerous transients
found in an optical survey triggered by the neutrino detection
requires extensive spectroscopic measurements, which may not always be feasible.
One of the practical solutions is to rely on the photometric redshift
of the host galaxies whose information is available in the data taken by
the survey facilities with the wide field of view
such as the Vera C. Rubin Observatory.
Another way to perform the intensive spectroscopy will be brought by
the prime focus spectrograph on Subaru~\cite{2015JATIS...1c5001S}.
It has a remarkable capability
of a wide-field simultaneous spectrography with high multiplicity.
It is of great importance for neutrino and optical astronomers to closely collaborate
for discoveries of yet unknown transient neutrino sources.

Searches for neutrino multiplets are powerful not just for reliably identifying the EMG counterparts,
but also for revealing/constraining the source characteristics.
Figure~\ref{fig:number_of_multiplet_s} shows the number of sources to yield
detectable neutrino multiples, $N^{\rm M}_{\Delta \Omega}$,  from the sky patch of 
$\Delta \Omega= 1\ {\rm deg}^2$~\cite{Yoshida:2022idr}. The search time scale is assumed to be
$\Delta T = 30$~days, considering the typical time scale of the neutrino transient emission
from TDEs and core-collapse SNe and being long enough to cover faster transients such as
low luminosity GRBs and GRB afterglow.
Only the diagonal region is displayed where
it is consistent with the cosmic neutrino diffuse background flux,
$E_\nu^2\Phi_{\nu_e+\nu_\mu+\nu_\tau}\approx 10^{-8}\ {\rm GeV\ cm^{-2}\ sec^{-1} sr^{-1}}$.
Because a neutrino telescope looks for $\sim 2\pi$ sky,
the parameter space of $N^{\rm M}_{\Delta \Omega} \gtrsim 10^{-6}$ is accessible
by a $1\ {\rm km}^3$ neutrino telescope with 5 year observation.
Null detection of the neutrino multiplets with the criteria
of the false alarm rate ${\rm FAR}\lesssim 0.25\ {\rm yr}^{-1}$
leads to the allowed parameter space of
\begin{equation}
    {\mathcal E}_\nu^{\rm fl} \lesssim  5\times 10^{51}\ {\rm erg}, \quad
    R_0  \gtrsim  2\times 10^{-8} ~{\rm Mpc}^{-3}~{\rm yr}^{-1},
    \label{eq:constraints_on_source_parameters}
\end{equation}
if the transients of $\Delta T\lesssim 30$~days are the major sources
to contribute the high energy cosmic neutrino diffuse background radiation~\cite{Yoshida:2022idr}.
It will constrain the models involving jetted TDEs and super-luminous SNe
in future.

\section{Summary}
The observational fact that the neutrino energy flux at $100\ {\rm TeV}\sim 1\ {\rm PeV}$
is comparable with that of UHECRs is understandable by the UHECR-neutrino multi-messenger
approach. Based upon the framework of photo-hadronic interactions for producing the secondary neutrinos,
we have constructed the generic scheme to describe the common sources
of UHECRs and high energy neutrinos with less model-dependent way and
obtained the viable parameter space required to explain the diffuse high-energy neutrino flux above
100 TeV energies and the UHECR flux above 10 EeV, simultaneously.
The five conditions have essentially constrained the allowed parameter space which is rather
narrow -- requirements for the UHECR acceleration, the UHECR escape, the UHECR nuclei survival,
the UHECR energetics, and the fiducial neutrino flux.
For an astronomical object with a given photon luminosity, a number density,
magnetic field strength, and its equipartition parameter $\xi_{\rm B}$,
the basic requirements to qualify as the common sources are represented in
the form of conditions of $\xi_{\rm B}$, $p\gamma$ optical depth at UHECR energy
of 10 PeV, and the CR loading factor $\xi_{\rm CR}$.
Among the known astronomical object classes,
we have found that the low luminosity GRBs are the most likely candidate,
followed by the jetted TDEs and FSRQs with the extreme parameter tuning.
We would like to emphasize, however, that our framework to provide the generic constraints
is applicable to as-yet-unknown source populations which may be revealed in
future neutrino-driven multi-messenger observations.

These common source candidates are transients in optical and X-ray bands.
In addition, many other neutrino source candidates at TeV sky such as the circumstellar SNe
are also transients. Thus it is a key to conduct the multi-messenger follow-up
observations in oder to identify the neutrino sources. Overcoming the difficulty
that the optical transient sky is filled with numerous SNe, the two approaches have been proposed
with the presently operating facilities --
introducing a sub-threshold detection channel to a wide field of view X-ray observatory
responding to high energy neutrino detection, and the neutrino multiplet detection
to trigger ToO observations with optical telescopes.
We are living in a new era to utilize neutrino, optical, and X-ray messengers
to reveal the origin of cosmic rays, and study the hadronic emissions.

\bigskip
\textbf{Acknowledgments}

\vspace{0.2cm}
The studies written in this article have benefited from
the extensive discussions with Kohta Murase. The author is also
deeply grateful to Masaomi Tanaka, Nobuhiro Shimizu, and Aya Ishihara
for their valuable inputs. Special thanks go to
IceCube Collaboration. This work is supported partly by
JSPS KAKENHI grant No.18H05206 and 23H04892.

\bibliography{syoshida}
\bibliographystyle{unsrt}

%
%
%

\end{document}